\documentclass[prl,aps,showpacs,twocolumn,superscriptaddress]{revtex4-1}
\usepackage{amsmath,amssymb,graphicx}
\usepackage{amsmath}

\begin{document}

\title{On-chip optical event horizon}

\author{Charles Ciret}
\email{charles.ciret@ulb.ac.be}
\affiliation{OPERA-Photonique, Universit\'e libre de Bruxelles (ULB), 50 av. F.D. Roosevelt, CP194/5, B-1050 Bruxelles, Belgium}
\author{Fran\c{c}ois Leo}
\affiliation{OPERA-Photonique, Université libre de Bruxelles (ULB), 50 av. F.D. Roosevelt, CP194/5, B-1050 Bruxelles, Belgium}
\affiliation{Photonics Research Group, Department of Information Technology, Ghent University-IMEC, Ghent B-9000, Belgium}
\affiliation{Center for Nano- and Biophotonics (NB-photonics), Ghent University, Ghent B-9000, Belgium}
\author{Bart Kuyken}
\author{Gunther Roelkens}
\affiliation{Photonics Research Group, Department of Information Technology, Ghent University-IMEC, Ghent B-9000, Belgium}
\affiliation{Center for Nano- and Biophotonics (NB-photonics), Ghent University, Ghent B-9000, Belgium}
\author{Simon-Pierre Gorza}
\affiliation{OPERA-Photonique, Université libre de Bruxelles (ULB), 50 av. F.D. Roosevelt, CP194/5, B-1050 Bruxelles, Belgium}




\begin{abstract}
The interaction of waves in nonlinear Kerr waveguides are, under some circumstances, similar to the physics occurring at the horizon of black and white holes. Here, we investigate this analogy in an integrated nonlinear photonic structure through the reflection of a continuous wave on an intense pulse. The resulting frequency conversion of the continuous wave is experimentally demonstrated in a silicon photonic wire waveguide and confirmed by simulations using the generalized nonlinear Schrödinger equation. This demonstration paves the way to efficient all-optical signal processing functionality integrated on a CMOS-compatible waveguide platform.
\end{abstract}


\maketitle


The manipulation of optical signals by light itself through the interaction of pulses in nonlinear media is an active research area. The goal is to develop low power consumption ultra-fast all-optical devices, for replacing electronics-based systems used today. In this respect, the realization of an all-optical transistor has attracted a lot of attention from the scientific community over the past decades, and still constitutes an active area of research \cite{Miller:NatPhotonics:2010,Demircan:11:Phys.Rev.Lett}. However, mandatory arduous criteria have to be fulfilled for its demonstration \cite{Miller:NatPhotonics:2010}, in particular fan-out and cascadability, which are the most difficult to meet. Recently, it has been shown that such devices, with at their heart the strong light-light interaction arising at an optical event horizon, can be realized  \cite{Demircan:11:Phys.Rev.Lett,Tartara:15:JOSAB}.

The optical analogue of an event horizon occurs when an intense pulse, propagating in a nonlinear waveguide, prevents a weak probe wave traveling at a different velocity from passing through it. This analogy is both interesting for fundamental science as for practical applications. It has been shown that an optical event horizon can mimics the effect of a black or white hole. As such it can act as a laboratory scale universe, to research effects such as the Hawking radiation \cite{Philbin:08:Science,Hawking:74:Nature}. Moreover, because of the inherent frequency shift that takes place during the interaction between the two waves, the phenomenon has been pointed out as a mechanism for efficiently converting the frequency of a signal \cite{Tartara:12:IEEE.J.Quantum.Electon.}. Note that, the Cherenkov radiation, also known as dispersive wave generation in the extensively studied context of supercontinuum generation, constitutes also an efficient way for the transfer of energy far from the pump wavelength \cite{Dudley:06:Rev.Mod.Phys}.    

The first demonstrations of optical event horizons have been realized in optical fibers, by superimposing at the fiber input a weak probe wave together with an intense pulse \cite{Philbin:08:Science,Webb:14:Nature.Commun}. More recently, they have been studied through the collision between pulses, generated in topographic fibers designed for that purpose \cite{Bendahmane:15:OptExpress}. In these experiments, the weak nonlinearity of silica results in the need for pulses with kilowatt peak power and long interaction lengths, clearly preventing the use of optical fibers for the aforementioned integrated low power applications. Nonlinear interactions in silicon nanophotonic waveguides, on the contrary, arise for watt-level peak power pulses and for millimeter-scale propagation lengths. Such a platform has thus successfully been used, amongst other, for supercontinuum generation at telecommunication wavelengths in millimeter size optical chips \cite{Zhang:12:IEEE,Halir:12:Opt.Lett,Leo:14:Opt.Lett}. In addition, these nanophotonic structures are fully CMOS-compatible making them very good candidates for the realization of integrated low power optical functionalities in a, low-cost, high volume fabrication platform. In this letter, we experimentally demonstrate, to the best of our knowledge the first optical event horizon in an integrated photonic nonlinear structure through the reflection of a weak continuous probe wave on an intense pump pulse. These results pave the way to all-optical functionalities based on the optical analogue of an event horizon in  a CMOS-compatible platform.

\begin{figure}[t!]
\centering
\includegraphics[width=\linewidth]{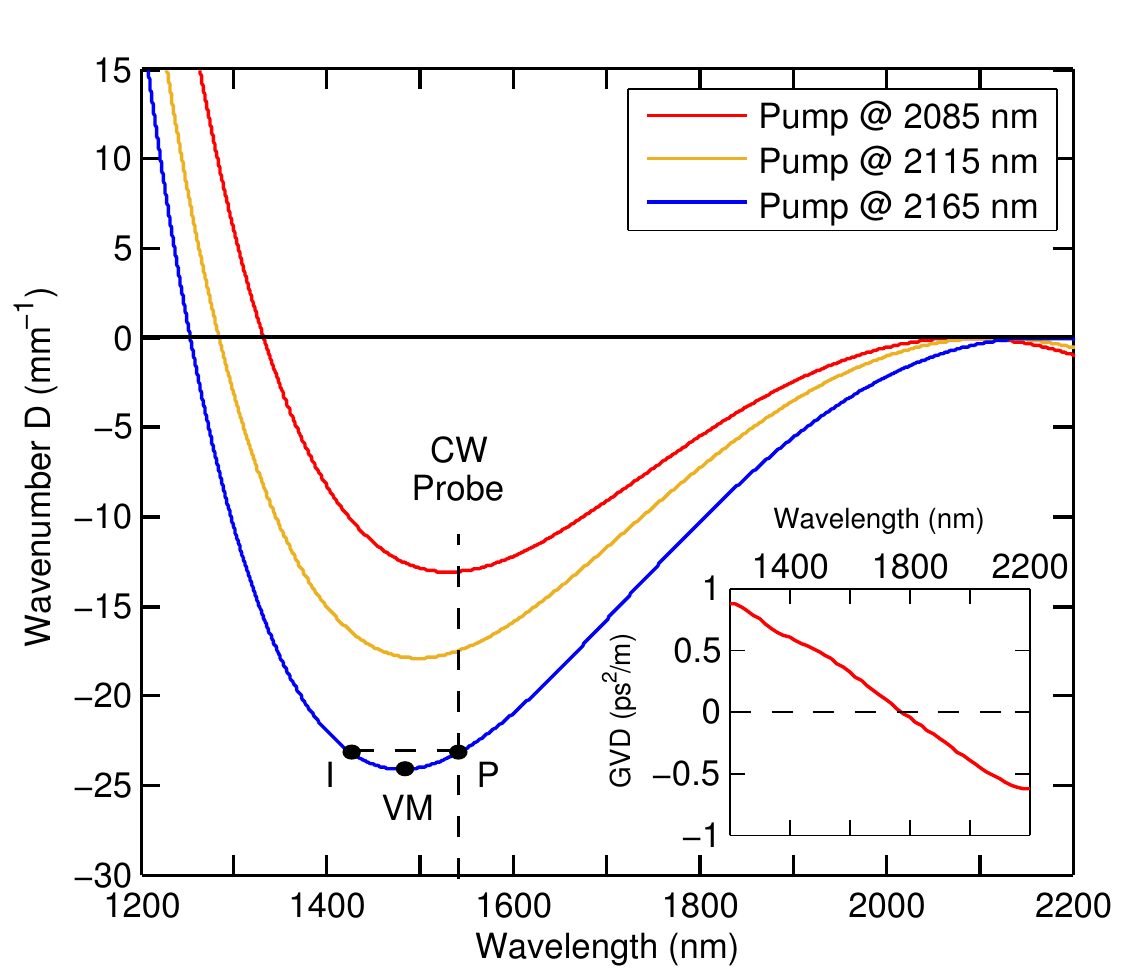}
\caption{Wavenumber $D$ as a function of the wavelength for different pump wavelengths. The vertical dashed line represents the position of the continuous probe wave at 1541~nm. The GVD curve for a 220~nm-thick, 880~nm-wide silicon nanophotonic waveguide is shown in the inset. GVD: group velocity dispersion, CW: continuous wave. P, I and VM points, see discussion in the text (indicated only for the blue curve for clarity). }
\label{Wavenumber}
\end{figure}

Our demonstration is realized in a 22~mm-long silicon-on-insulator waveguide with a height of 220~nm and a width of 880~nm. This waveguide has been carefully chosen for its linear dispersion properties. It possesses a zero-dispersion wavelength at 1740~nm (see in the inset of Fig.~\ref{Wavenumber}), located between the telecom C-band where the probe is positioned and the pump wavelength positioned around 2~$\mu$ m in the anomalous dispersion regime.


When propagating in the waveguide, the nonlinear interaction between a continuous probe wave (CW probe) at $\omega_{probe}$ and a pump wave at $\omega_{pump}$ can be interpreted as a cascaded-four wave mixing mechanism, leading to a frequency conversion of the CW probe. It has been shown that the conversion from the probe wavelength to a single higher-order idler wavelength component $\omega_{idler}$ is efficient when the resonant condition \cite{Webb:14:Nature.Commun,Xu:13:OptLetters}:

\begin{equation}
D(\omega_{idler}-\omega_{pump})=D(\omega_{probe}-\omega_{pump}),
\label{resonant_condition}
\end{equation} 
is fulfilled. The wavenumber $D(\omega-\omega_{pump})=\beta(\omega)-\beta_0-\beta_1 \times  (\omega - \omega_{pump} )$ [where $\beta_0=\beta(\omega_{pump})$ and $\beta_1=d\beta/d\omega|_{\omega_{pump}}$] is the wavenumber of a linear wave at a frequency $\omega$ in a reference frame co-moving with the pump pulse at a frequency $\omega_{pump}$. In the temporal domain, this frequency shift is the result of the nonlinear phase imprinted upon the CW wave by the pump pulse.    

The evolution of the wavenumber $D$ as a function of wavelength for different pump wavelengths is displayed in Fig.~\ref{Wavenumber} for the waveguide used in the experiment. Considering a CW probe at 1541~nm (see dashed line as well as the P point on the blue curve), we can infer from this figure that the idler wave satisfying the resonance condition Eq.~(\ref{resonant_condition}) (see the I point for the blue curve) is located on the other side of the minimum of the wavenumber $D$ versus wavelength curve (VM point for the blue curve).  
This peculiar position corresponds to the wavelength that is group velocity matched with the pump wave. For a 2165~nm pump (blue curve), the CW probe at 1541~nm travels faster than the pump and the change in the group velocity resulting from the frequency shift prevents the CW wave to cross the pump pulse. The trailing edge of the pump pulse thus corresponds to an optical horizon for the probe, similar to a white-hole horizon. A similar reasoning can be made for a probe traveling slower than the pump, in this case the leading edge of the pump is similar to a black-hole horizon.

\begin{figure}[t!]
\centering
\includegraphics[width=0.9\linewidth]{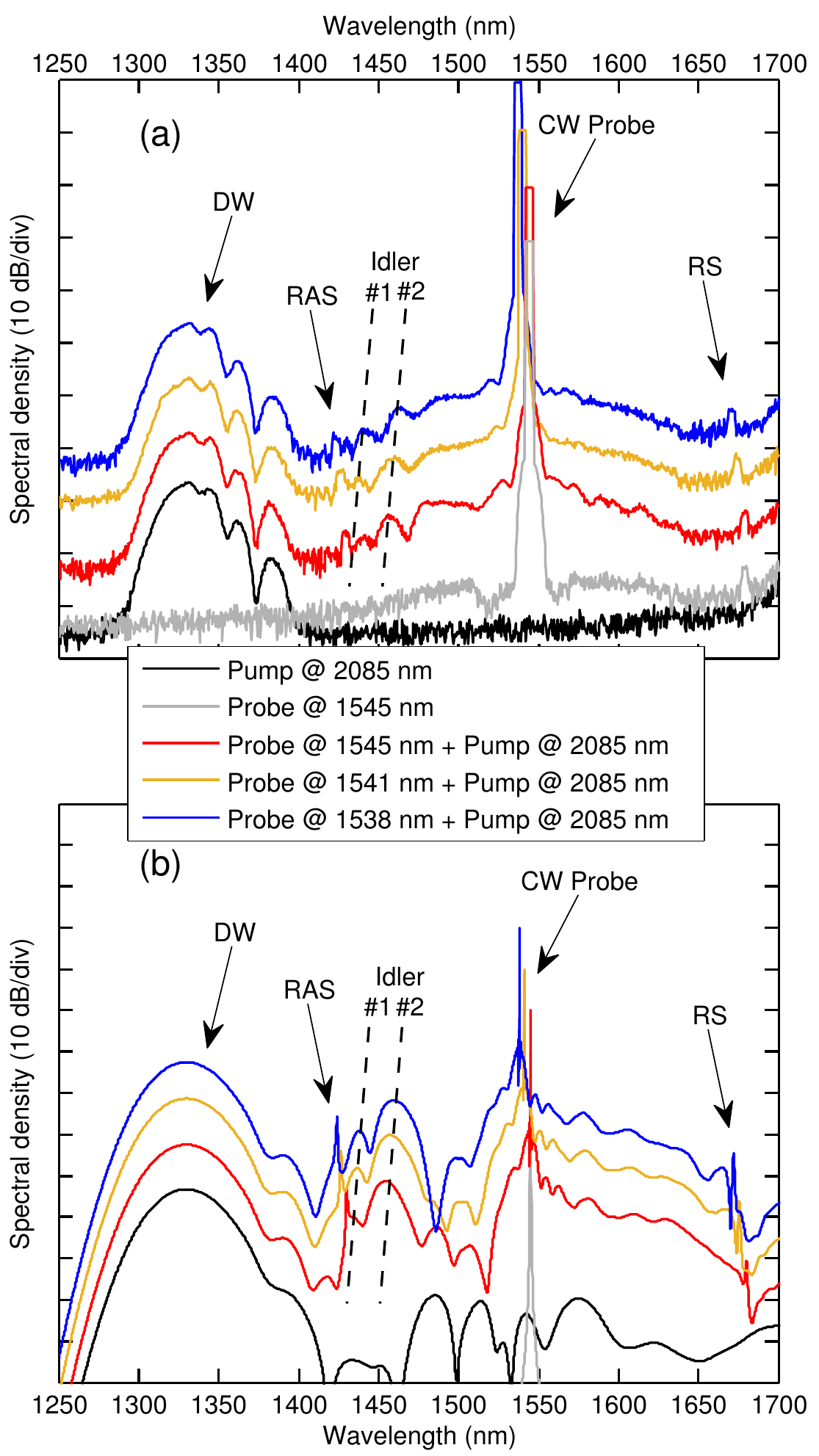}
\caption{Experimental spectra recorded at the output of our 22~mm-long Si waveguide (a) and the corresponding simulated spectra (b) for a 6.5~W, 180~fs \textit{sech}$^2$-shaped input pump wave at 2085~nm. These curves show the output spectra with the pump alone (black curves), with only the 500~$\mu$W CW probe at 1545~nm (gray curves), and with both the pump and the CW probe at different wavelengths (other colors). The dashed lines show two reflected waves, that are shifted when the probe wavelength is tuned. The curves are shifted by 10 dB for clarity, except for the gray and the black curves. DW: dispersive wave, CW: probe wave, RAS: Raman anti-Stokes, RS: Raman Stokes.}
\label{Exp_Simu}
\end{figure}

In the experiments, the intense pump pulse at a wavelength of 2085~nm is generated  by an optical parametric oscillator (OPO, Spectra Physics OPAL) delivering 180~fs pulses (full width at half maximum, FWHM) at 82~MHz repetition rate. The CW probe, for which the wavelength can be tuned from 1538~nm to 1545~nm, is generated by a CW laser followed by an Er-doped fiber amplifier. At the amplifier output, the CW beam is collimated and the two beams are combined by means of a dichroic mirror. The beams are then coupled into the photonic nanowire using a microscope objective, each of them exciting only the quasi-TE mode of the waveguide. At the wire output, the light is collected by a lensed fiber and sent to an optical spectrum analyzer (OSA). The experimental output spectra recorded in the range 1250~nm-1700~nm, i.e. up to the wavelength cut-off of our OSA, are displayed in Fig. \ref{Exp_Simu}(a) for an on-chip peak pump power of  6.5~W and a 500~$\mu$W CW probe. In the experiment, the pump power was progressively increased until the reflection of the probe was clearly visible in the spectrum. The corresponding simulations are shown in Fig. \ref{Exp_Simu}(b). These simulations were performed by solving the generalized nonlinear Schrödinger equation (GNLSE), where higher-order chromatic dispersion terms and the first order dispersion of the nonlinearity are included:

\begin{multline}
\frac{\partial A(z,t)}{\partial z}=i\sum_{k=2}^7\frac{i^k}{k!}\beta_k\frac{\partial^kA(z,t)}{\partial t^k}-\frac{\alpha_0}{2}A(z,t) \\ +i\gamma(1+\frac{i}{\omega_{pump}}\frac{\partial}{\partial t})A(z,t)\int_{-\infty}^{+\infty} R(t')\left| A(z,t-t')\right|^2 dt'.\label{EqGNLS}
\end{multline}
In this equation, $A(z,t)$ is the slow-varying envelope of the electric field, $\beta_k$ are the dispersion coefficients associated with the Taylor series expansion of the propagation constant around the pump frequency $\omega_{pump}$, $\alpha_0=2 \text{dB}/\text{cm}$ is the linear loss coefficient, $\gamma=(234+i15)\text{ W}^{-1}\text{m}^{-1}$ is the complex nonlinear parameter and $R(t')$ is the response function accounting for the instantaneous and delayed Raman contributions to the nonlinearity. The effects of free carriers are not included in this model because we have numerically verified that their contributions are negligible in our experimental configuration.  

\begin{figure}[t!]
\centering
\includegraphics[width=\linewidth]{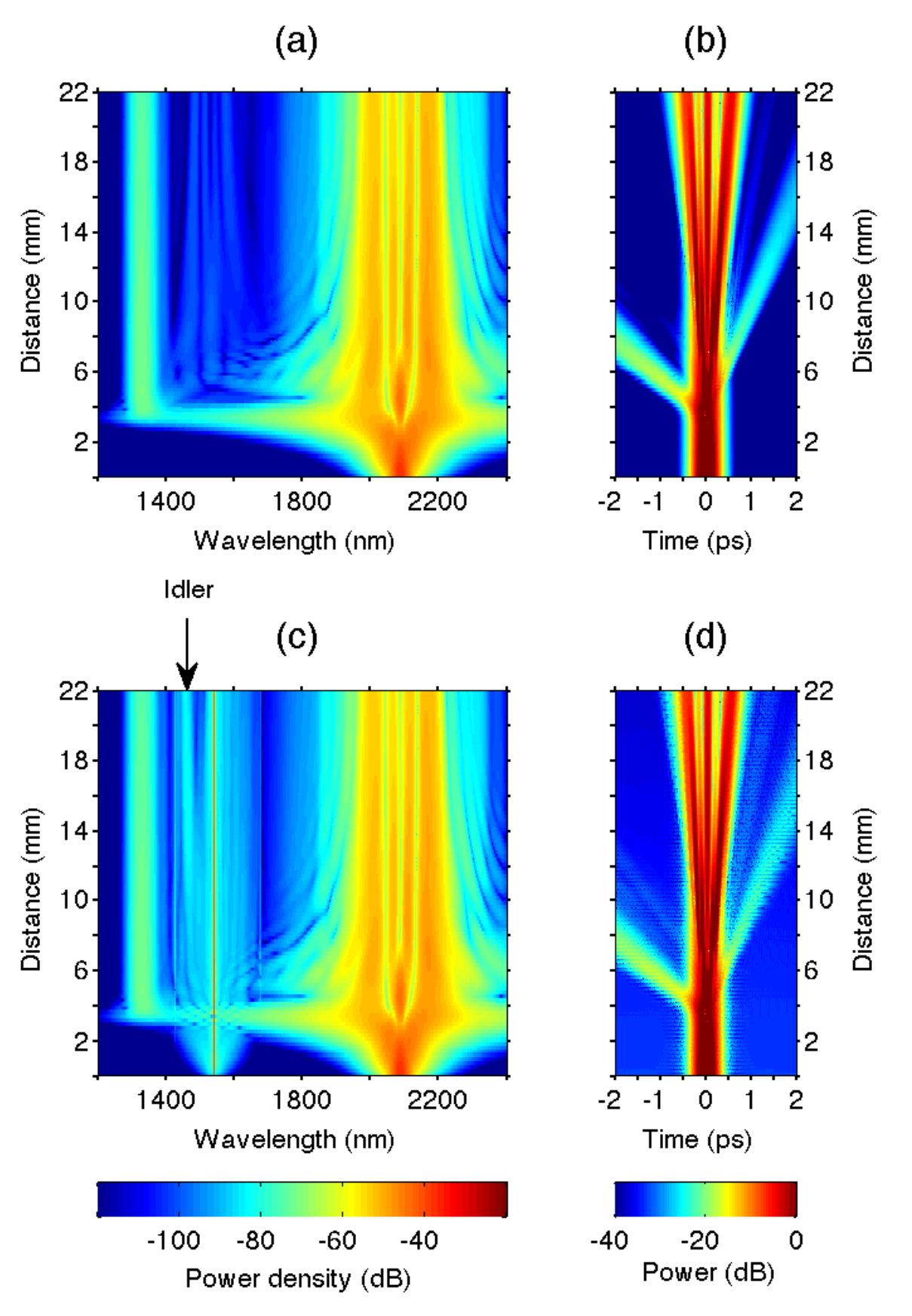}
\caption{Pseudocolor plots of the simulated spectral [(a), (c)] and temporal [(b), (d)] evolutions along the 22~mm propagation for a 6.5~W, 180~fs \textit{sech}$^2$-shaped input pump wave at 2085~nm propagating without [(a), (b)] and with the 500~$\mu$ W CW probe at 1541~nm [(c), (d)]. The arrow highlights the frequency conversion of the CW probe toward the two idler waves at 1457~nm and 1436~nm. Note that in this figure the two idler waves are not spectrally resolved.}
\label{Simu_Propa}
\end{figure}

\begin{figure}[t!]
\centering
\includegraphics[width=\linewidth]{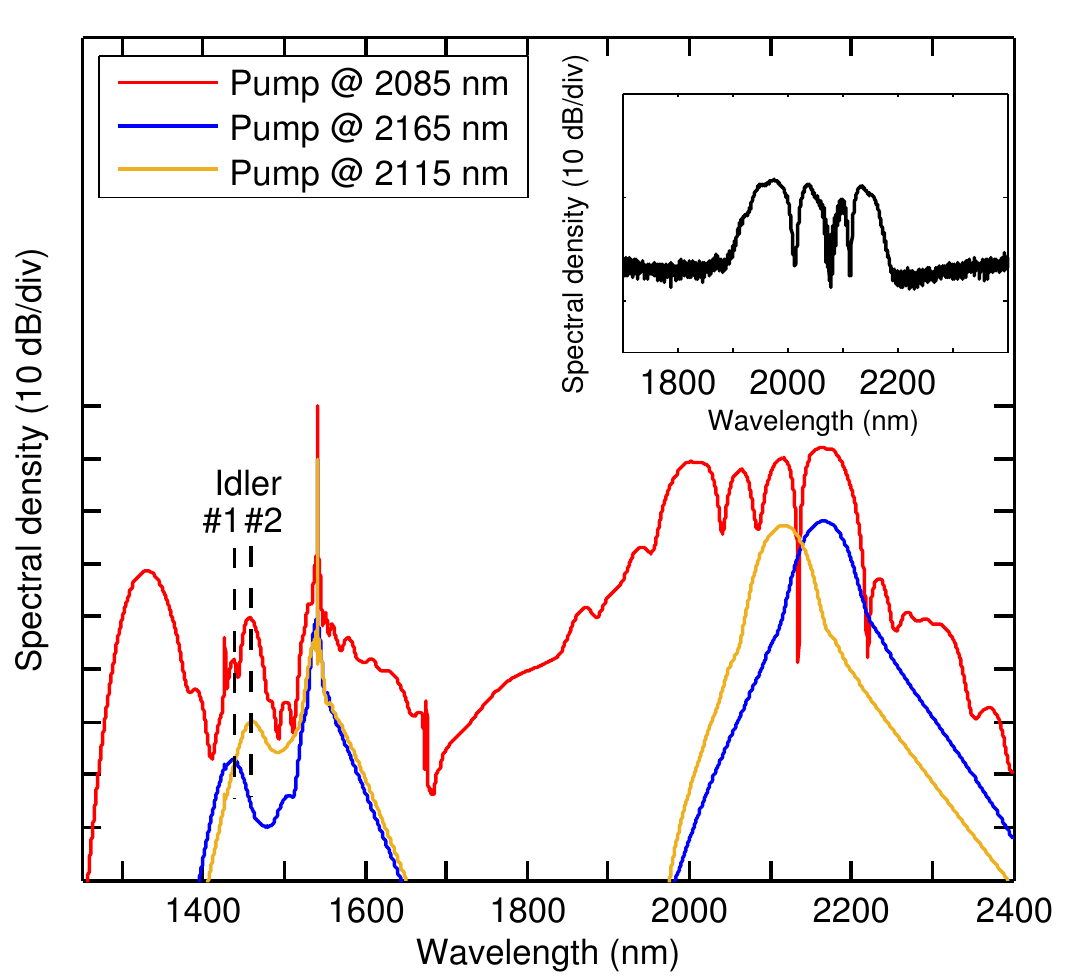}
\caption{Simulated output spectra for a 500~$\mu$ W CW input probe at 1541~nm together with a 6.5~W, 180~fs \textit{sech}$^2$-shaped input pump wave at 2085 nm (red curve) or with a 1.8~W, 70~fs \textit{sech}$^2$-shaped input pump wave at 2115~nm (yellow curve) or at 2165~nm (blue curve). The corresponding output spectrum, recorded with a Fourier transform optical spectrum analyser, for a 6.5~W, 180~fs input pump wave at 2085~nm is shown in the inset in the wavelength range 1700~nm-2400~nm. The red curve is shifted by 10~dB for clarity.}
\label{Pump70fs}
\end{figure}

At the input peak power needed to observe the reflection, it can be seen in Fig.~\ref{Exp_Simu} (black curves) and Figs.~\ref{Simu_Propa}(a,b) and ~\ref{Pump70fs}, that the pump pulse propagation in the anomalous dispersion region leads to the generation of a supercontinuum extending from 1.3\,$\mu$m to 2.2\,$\mu$m. As shown previously, the spectral broadening is the result of self-phase modulation, soliton splitting, as well as the emission of a dispersive wave (DW) \cite{Yin:07:Opt.Letters,Leo:14:Opt.Lett}. The DW is generated in the normal dispersion region, around 1330~nm. This wavelength corresponds to the degenerate case in the resonant condition Eq.~(\ref{resonant_condition}) where $\omega_{probe}=\omega_{pump}$. It is thus given by the crossing points between the $D$ curves and the horizontal black line in Fig.~\ref{Wavenumber}. When only the pump pulse is coupled in the waveguide, only the DW is visible in the recorded output spectrum shown in Fig.~\ref{Exp_Simu}(a). The output spectrum for the CW alone is shown as the gray curves in Fig.~\ref{Exp_Simu} and is essentially undistorted because of the low CW power and the normal dispersion.

However, when both the pump and the CW probe are propagating together, the output spectra show other distinctive features. Close to the probe wavelength a wide pedestal can be seen. It is generated during the first part of the propagation dynamics [see Fig.~\ref{Simu_Propa}(c)] and is the result of the cross phase modulation on the CW probe which overlaps with the input pump pulse. On either sides of the probe, two narrow peaks, labeled RS and RAS in Fig.~\ref{Exp_Simu}, can also be identified. These peaks are $\pm$17~THz away from the wavelength of the CW probe and correspond to the generation of Raman anti-Stokes (RAS) and Raman Stokes (RS) photons from the probe wave, as confirmed by canceling in simulation the delayed nonlinear term in Eq.~(\ref{EqGNLS}). The remaining features are the two broad peaks labeled Idler $\#1$ and $\#2$. Their positions agree quite well with the numerical simulations and are clearly shifted toward the probe  for decreasing probe wavelength. The simulated spectral and temporal evolution along the propagation distance plotted in Fig.~\ref{Simu_Propa}(c)-(d) show that the reflection of the CW probe actually occurs after the soliton fission that follows the temporal pulse compression [Note that, as expected and contrary to ~\cite{Demircan:11:Phys.Rev.Lett,Webb:14:Nature.Commun}, the reflection of the probe onto the pump is not clearly visible in the simulated temporal evolution plotted in Fig.~\ref{Simu_Propa}(d) because the probe is a continuous wave giving an almost temporal uniform background]. The fundamental solitons resulting from the fission have a pulse duration of 70~fs  and are centered around 2115~nm, 2165~nm and 1995~nm. The simulations displayed in Fig.~\ref{Pump70fs} with an input pulse corresponding to one of these fundamental solitons with the same peak power as just after the fission (1.8~W) clearly show that, in our experiment, the Idler $\#1$ wave is the result of the reflection of the probe on the soliton emitted at 2165~nm and the Idler $\#2$ is related to the reflection on the soliton at 2115~nm. The idler wavelengths at 1436~nm and 1457~nm, respectively, are in excellent agreement with the resonant condition Eq.~(\ref{resonant_condition}) as displayed in Fig.~\ref{Wavenumber}.     
This confirms the central role played by the resonant condition on the wavenumber $D$ as seen in the previous demonstrations of optical event horizons in optical fibers \cite{Webb:14:Nature.Commun}. Along the same lines, we have verified that the idler wavelength depends on the probe wavelength. Both experimentally and numerically, we can observe in Fig.~\ref{Exp_Simu} that when the CW probe is shifted toward the zero slope wavelength, the idler wavelength also gets closer to that wavelength in accordance with Fig.~\ref{Wavenumber}.
Note that since a CW probe has been used in this proof of principle experimental work, only a very small part of the probe interacts with the pump pulse train. The frequency conversion is thus not expected to be efficient. However, by replacing the CW probe by a pulsed probe, the frequency conversion can be very efficient as previously shown \cite{Tartara:12:IEEE.J.Quantum.Electon.}. For non negligible frequency conversion, the resulting spectral change on the soliton pump pulse affects its velocity, a mechanism that has been recognized to be suitable to make useful optical transistors \cite{Demircan:11:Phys.Rev.Lett, Miller:NatPhotonics:2010}.

In conclusion, we have experimentally studied an optical analogue of an event horizon in a silicon on insulator nanophotonic waveguide. We have unambiguously observed the frequency conversion of a CW wave when propagating with an intense pump pulse. This conversion occurs when its wavelength is close to the wavelength that is group velocity matched with the pump. The excellent agreement between the GNLS modeling and the experimental results reveals that in our experiment the reflection takes place after the soliton fission. An even better demonstration is thus in principle possible by resorting to shorter pump pulses of the order of 50\,fs. Our experimental results improve the understanding of nonlinear interactions in silicon waveguides with ultrashort pulses and foster further study of the potential of on-chip efficient frequency converters and all-optical transistors based on the physics of the optical analogue of an event horizon.

\section*{Funding Information}
This work was supported by the Belgian Science Policy Office (BELSPO) Interuniversity Attraction Pole (IAP) project Photonics@be and by the Fonds de la Recherche Fondamentale Collective, Grant No. PDR.T.1084.15.

%
%
%






\begin{thebibliography}{10}
\newcommand{\enquote}[1]{``#1''}

\bibitem{Miller:NatPhotonics:2010}
D.~A.~B. Miller, \enquote{Are optical transistors the logical next step?} Nat
  Photon \textbf{4}, 3--5 (2010).

\bibitem{Demircan:11:Phys.Rev.Lett}
A.~Demircan, S.~Amiranashvili, and G.~Steinmeyer, \enquote{Controlling light by
  light with an optical event horizon,} Phys. Rev. Lett. \textbf{106}, 163901
  (2011).

\bibitem{Tartara:15:JOSAB}
L.~Tartara, \enquote{Soliton control by a weak dispersive pulse,} J. Opt. Soc.
  Am. B \textbf{32}, 395--399 (2015).

\bibitem{Philbin:08:Science}
T.~G. Philbin, C.~Kuklewicz, S.~Robertson, S.~Hill, F.~K{\"o}nig, and
  U.~Leonhardt, \enquote{Fiber-optical analog of the event horizon,} Science
  \textbf{319}, 1367--1370 (2008).

\bibitem{Hawking:74:Nature}
S.~M. Hawking, \enquote{Black holes explosions?} Nature \textbf{248}, 30
  (1974).

\bibitem{Tartara:12:IEEE.J.Quantum.Electon.}
L.~Tartara, \enquote{Frequency shifting of femtosecond pulses by reflection at
  solitons,} Quantum Electronics, IEEE Journal of \textbf{48}, 1439--1442
  (2012).

\bibitem{Dudley:06:Rev.Mod.Phys}
J.~M. Dudley, G.~Genty, and S.~Coen, \enquote{Supercontinuum generation in
  photonic crystal fiber,} Rev. Mod. Phys. \textbf{78}, 1135--1184 (2006).

\bibitem{Webb:14:Nature.Commun}
K.~E. Webb, M.~Erkintalo, Y.~Xu, N.~G.~R. Broderick, J.~M. Dudley, G.~Genty,
  and S.~G. Murdoch, \enquote{Nonlinear optics of fibre event horizons,} Nat
  Commun \textbf{5} (2014).

\bibitem{Bendahmane:15:OptExpress}
A.~Bendahmane, A.~Mussot, M.~Conforti, and A.~Kudlinski, \enquote{Observation
  of the stepwise blue shift of a dispersive wave preceding its trapping by a
  soliton,} Opt. Express \textbf{23}, 16595--16601 (2015).

\bibitem{Zhang:12:IEEE}
L.~Zhang, Q.~Lin, Y.~Yue, Y.~Yan, R.~Beausoleil, A.~Agarwal, L.~Kimerling,
  J.~Michel, and A.~Willner, \enquote{On-chip octave-spanning supercontinuum in
  nanostructured silicon waveguides using ultralow pulse energy,} Selected
  Topics in Quantum Electronics, IEEE Journal of \textbf{18}, 1799--1806
  (2012).

\bibitem{Halir:12:Opt.Lett}
R.~Halir, Y.~Okawachi, J.~S. Levy, M.~A. Foster, M.~Lipson, and A.~L. Gaeta,
  \enquote{Ultrabroadband supercontinuum generation in a cmos-compatible
  platform,} Opt. Lett. \textbf{37}, 1685--1687 (2012).

\bibitem{Leo:14:Opt.Lett}
F.~Leo, S.-P. Gorza, J.~Safioui, P.~Kockaert, S.~Coen, U.~Dave, B.~Kuyken, and
  G.~Roelkens, \enquote{Dispersive wave emission and supercontinuum generation
  in a silicon wire waveguide pumped around the 1550 nm telecommunication
  wavelength,} Opt. Lett. \textbf{39}, 3623--3626 (2014).

\bibitem{Xu:13:OptLetters}
Y.~Q. Xu, M.~Erkintalo, G.~Genty, and S.~G. Murdoch, \enquote{Cascaded bragg
  scattering in fiber optics,} Opt. Lett. \textbf{38}, 142--144 (2013).

\bibitem{Yin:07:Opt.Letters}
L.~Yin, Q.~Lin, and G.~P. Agrawal, \enquote{Soliton fission and supercontinuum
  generation in silicon waveguides,} Opt. Lett. \textbf{32}, 391--393 (2007).

\end{thebibliography}
\end{document}